\documentclass[journal=mamobx,manuscript=article]{achemso}

\usepackage[version=3]{mhchem} % Formula subscripts using \ce{}
\usepackage[]{bm}
\usepackage{amssymb}

\usepackage{xr}
\usepackage{hyperref}
\usepackage{cleveref}
\usepackage{siunitx}
\usepackage{upgreek}
\usepackage{placeins}
\usepackage{textgreek}
\DeclareSIUnit\angstrom{\text {Å}}
\newcommand{\rev}[1]{\textcolor{black}{#1}}

\usepackage[font=scriptsize
]{caption}

\SectionNumbersOn

\author{Baris E. Ugur}
\author{Michael A. Webb}
\email{mawebb@princeton.edu}
\affiliation[Princeton University]
{Department of Chemical and Biological Engineering, Princeton University, Princeton, NJ, 08540, USA}

\title{Persistent Free Volume Governs (Anti)plasticization in Chitosan-Water Mixtures}

\begin{document}
\emergencystretch 3em
\renewcommand*{\figureautorefname}{Figure}
\Crefname{figure}{Figure}{Figures}
\renewcommand*{\equationautorefname}{Equation}
\Crefname{equation}{eq}{eqs}


\begin{abstract}

Chitosan is a highly versatile and sustainable polymer with a broad range of potential biological and materials engineering applications. Despite its versatility, the native brittleness of chitosan limits its broader utilization. This limitation can be addressed by blending chitosan with small-molecule additives to modulate its thermomechanical properties. We employ molecular dynamics (MD) simulations to investigate the mechanism underlying antiplasticization followed by plasticization at increasing water content. Decomposition of the elastic moduli reveals a competition between weakened polymer-polymer interactions and enhanced polymer-water interactions, with their relative strengths governing the resulting properties. We introduce a simple model incorporating dynamically accessible free volume regions as a key driver of polymer mobility, effectively capturing the (anti)plasticization of elastic properties. We show that accessibility of free volume regions is enabled by connectivity of additive-accessible volume regions. This study provides new insights into the molecular interactions that dictate the properties of chitosan-water mixtures and may inform the rational design of chitosan-based materials and other hydrated biopolymers.
\end{abstract}

\section{Introduction}

Chitosan is the deacetylated derivative of chitin, which is the second most abundant biopolymer on Earth.\cite{Piekarska2023}
As a material, chitosan possesses many desirable properties, including high mechanical strength, biodegradability,\cite{Xu1996} biocompatibility,\cite{Lee2022} non-toxicity,\cite{Salsabila2025} and antimicrobial activity.\cite{Guarnieri2022}
It is thus a versatile component that is deployed in numerous applications, such as drug delivery,\cite{Garg2019} wound treatment,\cite{RajinikanthB2024} food packaging,\cite{Florez2022} and diverse agricultural uses.\cite{Sharif2018,Bertrand2024}
For example, chitosan-based materials can enhance oral bioavailability of insulin,\cite{Seyam2020} prevent wound infections\cite{Burkatovskaya2006}, promote healing,\cite{Le2023} enhance food shelf life,\cite{Tuesta2025} and replace harmful pesticides.\cite{Liu2023}
Chitosan is heavily targeted as a sustainable alternative to traditional synthetic materials,\cite{Olunusi2025,Maluin2020} as evidenced by an estimated global market size of \$6.8 billion in 2019 and an annual growth rate of 24.7\%.\cite{polym13193256}
Therefore, there is significant interest in understanding the properties of chitosan-based materials and how they may be controlled or manipulated for possible applications.

The natural rigidity\cite{Gartner2011} and brittleness\cite{Khankhuean2024} of chitosan, however, limits its use in applications that require resilience and flexibility.\cite{Coer2025}
These limitations can be addressed by blending chitosan with small-molecule additives.\cite{Duan2022,AlLami2025,DarieNița2021}
This process results in plasticization phenomena, which is typically characterized by lower modulus,\cite{Eslami2023} glass transition temperature,\cite{Shen1965} and melt viscosity,\cite{Chen2005} along with increased gas permeability\cite{Petropoulos1992}; the additives are thus called plasticizers.\cite{Roussenova2010}
At low concentrations of certain additives in polymer mixtures, the opposite effect, known as antiplasticization, may occur.\cite{Seow1999}
The resulting properties of mixtures strongly depend on specific polymer-additive interactions governed by their chemistry.\cite{Kinjo1973}
Therefore, understanding the mechanism underlying these interactions is crucial to utilize plasticization and antiplasticization to manipulate the thermomechanical properties of otherwise desirable polymeric materials.

Historically, plasticization has been elucidated with lubricity,\cite{Kirkpatrick1940} gel,\cite{Aiken1947} and free-volume theories.\cite{MARCILLA2012,Daniels2009}
These theories were originally introduced for vinyl polymers in the 1930-40s to help rationalize experimental trends in glass transition temperatures and have since been used as a framework to guide polymer system design.\cite{Aiken1947,Kirkpatrick1940,MARCILLA2012}
Lubricity theory suggests that plasticizers act as internal lubricants that reduce friction between polymer chains and enhance flexibility.
Formation of plasticizer-rich gliding planes allows polymer chains to glide over each other upon an applied force, increasing mobility.
While similar in principle, the gel theory concentrates on the evolution of interaction strengths governing the polymer mixture.
In this view, plasticizers disrupt intermolecular forces within the polymer network, reducing the resistance to deformation arising from polymer interactions.
According to free-volume theory, plasticizers impart additional fractional free volume within the polymer matrix.
This additional unoccupied volume allows greater chain segment mobility and enhances cooperative molecular motion, reducing glassiness and rigidity.
Studies on plasticized chitosan systems attribute plasticization to lubricity or gel theories, determined by the specific plasticizer chemistry.\cite{Suyatma2005,Mars2024,Smith2021}
As such, elucidating plasticization with these theories remains a challenge as these mechanisms are not generalizable to all polymer-additive systems, notably those exhibiting antiplasticization.\cite{Anderson1995}

Numerous studies experimentally probe (anti)plasticization behavior for diverse systems, including chitosan.
Plasticization of chitosan with glycerol has been attributed to the disruption of glucosamine intramolecular hydrogen bonds with insights from attenuated total reflectance-Fourier transform infrared spectroscopy (ATR-FTIR) and nuclear magnetic resonance (NMR).\cite{Smith2021}
However, the proposed mechanism does not account for the role of water despite the high moisture uptake of chitosan at ambient conditions.\cite{RemunanLopez1997,Correlo2007}
Furthermore, this mechanism focuses on the glucosamine monomer assemblies and may not translate to the chitosan polymer, where intermolecular interactions are expected to play a more significant role.
An electron spin resonance (ESR) study of chitosan-water mixtures reveals enhanced mobility in plasticized systems by measuring the rotational relaxation of a radical probe.\cite{MadeleinePerdrillat2016}
While this technique indicates enhanced plasticizer dynamics at the population level, it relies on indirect measurements using a radical probe and does not provide insights into the mechanism of plasticization.\cite{Plotnikov1977}
A positron annihilation lifetime spectroscopy (PALS) study\cite{Chaudhary2010} of chitosan-water systems reveals decreasing polymer free volume upon hydration, consistent with other hydrated biopolymers.\cite{Ubbink2016,Trotzig2009,Majstorovic2025}
While PALS provides sub-nanometer insights into the free volume trends in polymer-additive systems, the average hole size is obtained from a simplified model containing monodisperse spherical holes.\cite{Li1996,Tao1972}
However, it has been shown that these molecular voids are neither spherical nor monodisperse.\cite{Limbach2008}
Consequently, the atomic-level mechanism of (anti)plasticization of chitosan and other biopolymers is not yet fully understood.

Computational modeling methods can provide insight into the molecular-level physics that may not be accessible by experiment or theoretical frameworks.
Several studies employ MD simulations to characterize the thermomechanical properties of plasticized polymers, particularly biopolymers, across diverse chemistries.\cite{Chen2018,Oezeren2020,Oezeren2020a,Ugur2024}
However, there is very limited literature employing computational methods to elucidate the plasticization of chitosan systems.\cite{Chen2018}
Plasticization of starch and cellulose derivatives with additives, including water, polyols, and drug molecules, has been characterized with MD simulations.\cite{Oezeren2020,Oezeren2020a,Ugur2024}
An investigation of starch-glycerol systems has suggested the cooperative motion of local plasticized regions as the onset of effective plasticization.\cite{Oezeren2020}
In contrast, MD simulations of cannabidiol (CBD) amorphous solid dispersions containing cellulosic polymers have revealed that disruption of polymer-polymer hydrogen bonds by CBD molecules results in enhanced plasticization behavior.\cite{Ugur2024}
Nevertheless, even while plasticization is seemingly commonly observed, the underlying mechanisms are rarely scrutinized.
Moreover, there have been nascent reports of antiplasticization, highlighting another gap in our understanding.
As a result, existing models may not be generalized to other systems, particularly hydrated biopolymer systems spanning diverse chemistries.

In this study, we elucidate the nanoscale mechanisms underlying the composition-dependent (anti)plasticization behavior in chitosan–water mixtures using MD simulations.
We observe antiplasticization of thermomechanical properties at low water contents \rev{($<$10 wt.\%)} reflected by trends in the glass transition temperatures ($T_\text{g}$) and Young's modulus ($E$), consistent with experiment.
We report a rapid decrease in $E$ above 15 wt.\%, indicative of plasticization.
Decomposition of interactions contributing to material moduli reveals that the non-monotonic changes in $E$ are primarily governed by the competition between antiplasticizing polymer-water interactions and plasticizing polymer-polymer interactions.
We find that percolation of water clusters drives the crossover from antiplasticization to plasticization at higher additive contents.
We support our findings via detailed analysis of free volume, additive mobility, and hydrogen-bonding characteristics.
These molecular insights help understand the moisture-dependent behavior of diverse biopolymer systems and enable guided design of polymer-additive formulations with tailored properties. 
Our analysis framework can be applied to other polymer-additive systems, particularly those involving biopolymers, where different mechanisms may take part.

% \newpage

\section{Methods}

\subsection{Simulation}

\noindent\textbf{General Simulation Protocols.}
All MD simulations were carried out using the GPU-accelerated implementation of the OpenMM 7.7.0 simulation package.\cite{Eastman2017}
The systems were modeled with the Chemistry at Harvard Macromolecular Mechanics (CHARMM) force field\cite{Brooks2009}\rev{, which has extensive coverage for carbohydrates,\cite{R:2011_Guvench_CHARMM} and the TIP3P force field to describe the water molecules.}
\rev{For chitosan, we specifically use the parameters prescribed by Borca and Arango,\cite{Borca2016} who provided an extension of CHARMM to describe fully atomistic, cross-linked, and protonated chitosan models based on quantum chemical calculations.}
\rev{In this model, partial atomic charges were derived from Hartree–Fock/6-311G(d) single-point calculations on geometries pre-optimized using DFT at the B3LYP/6-311G(d) level, with electrostatic potential charge (ESP) fitting performed according to the Merz–Singh–Kollman scheme.}

Non-bonded interactions in real space were truncated at a distance of 10 \si{\angstrom}.
A switching function was applied between 9 and 10 \si{\angstrom} to gradually reduce non-bonded interaction energies to zero at the cutoff.
All 1-2 and 1-3 interactions were excluded from non-bonded calculations, while 1-4 interactions were included without scaling.
Long-range electrostatics were calculated using the Particle Mesh Ewald (PME) algorithm,\cite{PME} and energy corrections were applied for Lennard-Jones interactions beyond the cutoff range.
Isothermal-isobaric simulations used a timestep of 1 fs unless specified otherwise, a Langevin thermostat integrator with a friction coefficient of 1 ps\textsuperscript{-1}, and a Monte Carlo barostat with a frequency of 25 fs.

\noindent\textbf{System Preparation.}
\rev{For systems with 0, 5, 10, 15, 20, 30, 40 wt. \% water, a first set of four independent simulation replicates were prepared to assess statistical uncertainty.
Additional sets of four simulations were performed at 2, 4, 5, 6, 8 wt. \% water to better characterize the antiplasticization region.}
All mixtures were generated with 10 \rev{unprotonated chitosan} chains with a degree of polymerization of 50 and the minimum number of water molecules to reach the desired composition by weight.
\rev{A degree of polymerization of 50 was selected to ensure computational tractability while minimizing chain-end effects that manifest in shorter oligomers, consistent with chain lengths employed in previous MD studies of polymer mechanical properties.\cite{Elf2023,Nakamura2024,Oezeren2020}}
Each deacetylated chitosan chain was prepared with monomer units consisting of D-glucosamine rings linked together through $\beta$(1→4) glycosidic bonds.
Polymerization was performed with a maximum rotation angle of 30° between neighboring monomer units and a minimum contact distance of 1 \si{\angstrom} between constituent atoms.
The initial box size was chosen to obtain an initial mixture density of 0.1 g/cm\textsuperscript{3}.
The polymer chains and water molecules were subsequently distributed randomly within the simulation box with a random orientation.
The prepared systems were initially minimized using the OpenMM minimization algorithm with an energy tolerance of 10 kJ/mol, and initial velocities of particles were sampled from a Maxwell-Boltzmann distribution at 300 K.
The minimized structures were subsequently heated to their respective target temperatures ($T_\mathrm{target}$; see Supplemental Information, Table S1) at a rate of 200 K/ns, followed by a 21-step compression relaxation protocol (see Supplemental Information, Table S2) based on prior work \cite{Larsen2011}, all using a 0.5 fs timestep.
The output configurations were then annealed from their respective target temperatures to 200 K at a rate of 10 K/ns, incorporating an additional 2 ns of equilibration every 10 K within the isothermal-isobaric ensemble at 1 bar.
The 21-step compression relaxation protocol enables reliable convergence of material densities but not necessarily chain conformational statistics.\cite{Metcalfe2025}
\hfill

\noindent\textbf{Mechanical Properties.}
All mechanical analyses were performed using the last 1 ns of configurations from the equilibration run at 300 K.
Uniaxial and shear stress–strain responses were computed by incrementally deforming the simulation box by up to 10 \% and measuring the resulting stress tensors ($\sigma$) at the canonical ensemble.
A series of small strain ($\varepsilon$) perturbations were applied to the six unique components of the triclinic simulation box gradually over 100 steps, with 20 ps equilibration followed by 10 ps stress data collection per step, totaling 3 ns simulation time for each deformation direction.
The resulting stress and strain data was then used to compute the elasticity tensor:

\begin{equation} \label{elasticmatrix}
\mathbf{C} = \begin{pmatrix}
C_{11} & C_{12} & C_{13} & C_{14} & C_{15} & C_{16} \\
C_{12} & C_{22} & C_{23} & C_{24} & C_{25} & C_{26} \\
C_{13} & C_{23} & C_{33} & C_{34} & C_{35} & C_{36} \\
C_{14} & C_{24} & C_{34} & C_{44} & C_{45} & C_{46} \\
C_{15} & C_{25} & C_{35} & C_{45} & C_{55} & C_{56} \\
C_{16} & C_{26} & C_{36} & C_{46} & C_{56} & C_{66}
\end{pmatrix}
\end{equation}

\noindent where $C_{ij} = \sigma_i/\varepsilon_j$, and the indices $i$ and $j$ correspond to the stress and strain directions in the Voigt notation.
Each $C_{ij}$ was determined by fitting a linear regression to the initial, linear-elastic region of the corresponding stress–strain curve.
The computed elastic constants were then used to compute the mechanical properties:

\begin{equation}\label{bulk}
K = \frac{1}{3}(C_{11} + 2C_{12})
\end{equation}

\begin{equation}\label{shear}
G = C_{44}
\end{equation}

\begin{equation}\label{poisson}
\nu = \frac{1}{1 + \frac{C_{11}}{C_{12}}}
\end{equation}

\begin{equation}\label{elastic}
E = 2G(1 + \nu)
\end{equation}

\noindent where $K$ is the bulk modulus, $G$ is the shear modulus, $\nu$ is the Poisson's ratio, and $E$ is the Young's modulus of the material.

\noindent\textbf{Decomposition of Elastic Moduli Contributions}
Decomposition of contributions to Young's moduli was performed on a separate set of systems.
Configurations at 300 K were simulated for an additional 1 ns in the canonical ensemble.
Then, 2\% uniaxial strain was independently applied to the resulting configuration in $\mathrm{xx}$, $\mathrm{yy}$, and $\mathrm{zz}$ directions, and the system was equilibrated under each stress condition for 1 ns.
The last 0.5 ns data, corresponding to 50 configurations before and during each deformation were used to compute resulting changes in stress.
The respective changes in the stress, $\sigma_{\mathrm{xx}}$, $\sigma_{\mathrm{yy}}$, and $\sigma_{\mathrm{zz}}$ were used to calculate

\begin{equation} \label{uniaxial}
\Delta\sigma_\text{uniaxial} = (\Delta\sigma_\mathrm{xx}+\Delta\sigma_\mathrm{yy}+\Delta\sigma_\mathrm{zz})/3
\end{equation}

\noindent for $\varepsilon$ = 0.02.

\rev{Contributions to the stress from a specific interaction were calculated using a 
finite-difference approximation of the virial stress. 
For a given configuration 
with simulation cell vectors $\mathbf{h}$ and volume $V$, the cell is perturbed along 
$j$-th component of the $i$-th cell vector by a small displacement 
$\varepsilon L$, where we set the dimensionless strain magnitude as $\varepsilon = 10^{-4}$. 
Atomic positions are scaled 
affinely with the cell under each perturbation, and the potential energy 
$U_{kl}$ arising from interactions between species $k$ and $l$ is recomputed 
at the perturbed geometries. The stress contribution is then obtained via 
central finite difference:}

\begin{equation} \label{stress}
\rev{\sigma_{ij}^{kl} \approx \frac{1}{V} 
\frac{U_{kl}(\mathbf{h} + \varepsilon L \hat{\mathbf{e}}_{ij}) - 
      U_{kl}(\mathbf{h} - \varepsilon L \hat{\mathbf{e}}_{ij})}
     {2 \varepsilon }}
\end{equation}
\rev{where $ij$ denotes the component of the stress tensor, $k$ and $l$ 
denote the interacting species, $V$ is the simulation box volume, and 
$\hat{\mathbf{e}}_{ij}$ is the unit matrix with a single nonzero entry at 
position $(i,j)$. 
The total stress is recovered by summing over all pairwise 
contributions}
\begin{equation} \label{stress_sum}
\sigma_{ij} = \sum_{k=1}^N \sum_{l>k}^N \sigma_{ij}^{kl} + \sigma_{ij}^{\mathrm{bonded}}
\end{equation}
where $\sigma_{ij}^{\mathrm{bonded}}$ describes the stress 
contributions from bonded interactions, calculated \rev{via the same finite-difference 
procedure.}

    \begin{figure}[ht!]
\centering
\includegraphics[scale=1.0]{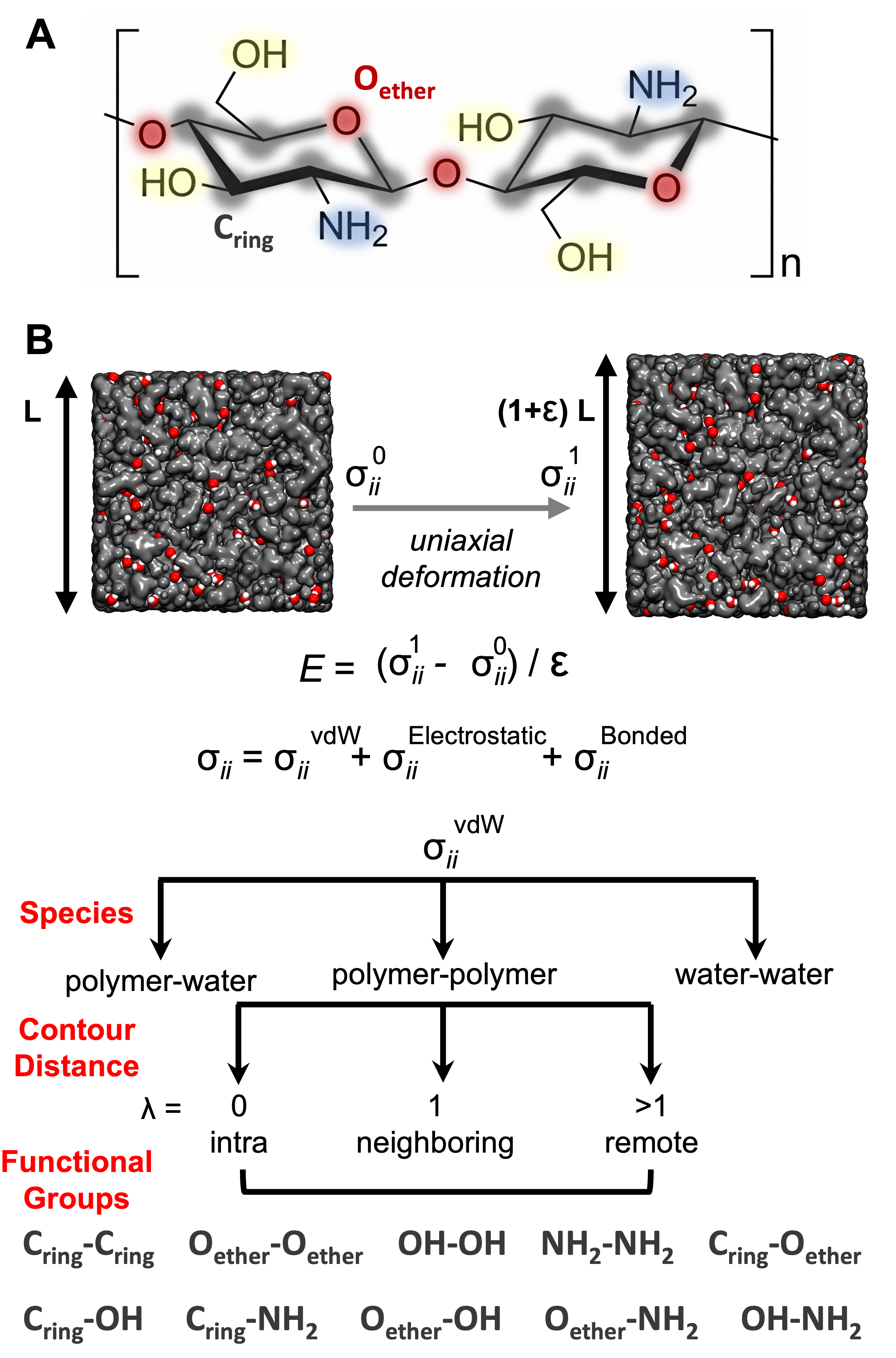}
\caption{Summary of the protocol for decomposing elastic moduli contributions. (A) Structure of chitosan containing D-glucosamine rings linked together through $\beta$(1→4) glycosidic bonds. Functional groups containing glucosamine ring carbons (gray, $\text{C}_\text{ring}$), ether oxygens (red, $\text{O}_\text{ether}$), hydroxyl groups (yellow, OH), and amine groups (blue, NH$_2$) are analyzed for their respective contributions to the elastic moduli.
(B) Visual representation of the uniaxial deformation protocol. Following deformation of the simulation box by $\varepsilon$ in the $ii$ direction, the subsequent change in the stress simulation stress contributions is used to compute the elastic modulus. These stress contributions are classified based on the interaction types (van der Waals, electrostatic, bonded). Stress contributions from van der Waals interactions are further classified into contributing species (polymer-water, polymer-polymer, water-water), and polymer-polymer interactions are decomposed based on the contour distance ($\lambda$=0, $\lambda$=1, $\lambda >$1) and specific functional groups.}
\label{fig1new}
\end{figure}

\hfill

\noindent\textbf{Glass Transition Temperatures.}
All simulated glass transition temperatures ($T_{\mathrm{g}}$) were obtained using simulated dilatometry by analyzing the temperature-dependent density ($\rho(T)$) data.
Densities averaged across the last 1 ns of equilibration simulations every 10 K were used for all $T_{\mathrm{g}}$ calculations.
The simulated $T_{\mathrm{g}}$ was identified as the temperature where the slope of $\rho(T)$ changes, signifying a shift in the thermal relaxation dynamics.

The simulated $T_{\mathrm{g}}$ were estimated by applying the broken-stick model to density-temperature data, with uncertainties quantified via a statistical bootstrapping procedure based on prior work.\cite{Deng2021,Chu2020}
The temperature dependence of densities in the glassy and rubbery regimes were determined using two temperature segments.
The segments were defined by their temperature bounds $T^\text{g}_\text{min}$ and $T^\text{g}_\text{max}$ (glassy), and $T^\text{r}_\text{min}$ and $T^\text{r}_\text{max}$ (rubbery).
The temperature bounds defining the segments for each system are listed in the Supplemental Information, Table S3. 
For each bootstrapping iteration, an endpoint ($T^\text{g}_\text{end}$ and $T^\text{r}_\text{end}$) was randomly chosen from each defined segment, and linear regressions were performed on the density data within [$T^\text{g}_\text{min}$, $T^\text{g}_\text{end}$] and [$T^\text{r}_\text{end}$,$T^\text{r}_\text{max}$].
The intersection temperature of the resulting lines was included as a $T_\mathrm{g}$ sample.
Intersection temperatures outside the range [$T^\text{g}_\text{end}$,$T^\text{r}_\text{end}$] were excluded from average $T_\mathrm{g}$ calculations.
This procedure was repeated until 10,000 samples were obtained, and the average intersection temperature was reported as the simulated $T_{\mathrm{g}}$ for the system.
Additional analysis of $T_\text{g}$ for each simulation can be found in the Supplemental Information (Figure S1).

\hfill

\noindent\textbf{Free-Volume Analysis.}
The simulation box was discretized into a three-dimensional grid, and spherical probes of 0.3 \si{\angstrom} radius were placed on grid points.
Any probe overlapping with an atom, defined as being within $\sigma/2$ of any atom where 
$\sigma$ is the Lennard–Jones diameter, was classified as occupied.
All unoccupied probes were considered free volume, and the free-volume fraction was calculated as $f_\mathrm{v} = V_\mathrm{free} / V_\mathrm{total}$.
The reported free-volume fractions were averaged over 100 frames collected throughout the last 1 ns of equilibration runs at 300 K.

To evaluate the contribution of each free volume grid point to particle mobility, time-averaged occupation probabilities $\bar{p}_\text{o}$ were calculated.
All grid points were classified using an indicator function $\textbf{1}_{o}(t)$ that specifies whether a point is occupied at time $t$ ($\textbf{1}_{o}(t)$=0 if free, $\textbf{1}_{o}(t)$=1 if occupied).
The indicator function was computed for all grid points across 100 frames from the last 1 ns of equilibration runs at 300 K.
For each grid point, the occupancy probability, $\bar{p}_\text{o}$, was computed as:

\begin{equation} \label{probability}
\bar{p}_\text{o} = \frac{1}{\Delta \tau} \int_{\tau _0}^{\tau _0 + \Delta \tau} \textbf{1}_{o}(\tau) d\tau
\end{equation}

\noindent
where $\Delta \tau$ is the occupancy interval length (0.5 ns), and $\tau _0$ is the initial frame (0.5 ns).
Free-volume grid point occupancy probability, $\bar{p}_\text{o,free}$ was computed only for free-volume grid points at the initial frame ($\textbf{1}_{o}(0)=0$). 

For water-accessible volume calculations, a 3 \si{\angstrom} diameter sphere was placed at the center of polymer free volume grid point, and the free-volume fraction within the sphere was calculated.
Grid points with a free-volume fraction higher than 50 \% were defined as water-accessible volume grid points.

\hfill

\noindent\textbf{Analysis of Dynamic Characteristics.}
For dynamic property analyses, configurations annealed to and subsequently equilibrated at 300 K were used.
Each system was further simulated in the canonical ensemble, using the Nosé–Hoover thermostat with a 1 ps$^{-1}$ collision frequency, for 0.5 ns for equilibration.
For short-time thermal and translational mobility analyses, the resulting systems were simulated for 100 ps where the configurations were saved every 20 fs.
To quantify local atomic fluctuations, Debye–Waller factors\cite{WidmerCooper2006} (DWFs) were computed for all water molecules atoms.

\begin{equation} \label{dwf_water}
\langle u^2 \rangle_{\text{water}} = \sum_{k}^{n_\text{w}}
\left (\frac{1}{3} \sum_{i \in k} \langle u^2_i \rangle \right)
\end{equation}

\noindent where $k$ and $i$ denote the indices of water molecules and their constituent atoms, and $n_\text{w}$ is the number of water molecules in the mixture.
All DWF analyses were performed over all configurations sampled throughout the 100 ps simulation.

To assess the short-term translational mobility within each mixture, the mean squared displacement (MSD) of water molecules were calculated via

\begin{equation} \label{msd}
\langle \Delta r^2(t) \rangle = \langle |\vec{r}(t) - \vec{r}(0)|^2 \rangle
\end{equation}

\noindent 
where $\vec{r}(0)$ and $\vec{r}(t)$ represent the initial position and the position at time $t$, respectively, of the center of mass of each water molecule.

\indent The orientational autocorrelation functions of water molecules were computed using:\cite{Yeh1999}

\begin{equation} \label{rot_autocorrelation}
C_{P_2}(\tau) = \left< P_2 [\textbf{u}(t_0) \cdot \textbf{u}(t_0 + \tau)] \right>
\end{equation}

\noindent
where $P_2$ denotes the second Legendre polynomial, while $\mathbf{u}$ denotes the unit vector along the average direction of the two O–H bond vectors.
The autocorrelation functions were fitted to a biexponential form, $A_1 e^{-t/\tau_1} + A_2 e^{-t/\tau_2}$, from which the rotational time constants were obtained as $\tau_{\mathrm{rot}} = A_1 \tau_1 + A_2 \tau_2$. 

\noindent\textbf{Water Cluster Analysis.}
Hydrogen-bonding analysis was performed to categorize cluster and isolated water molecules.
All hydrogen bonds in configurations from the 100 ps simulations in the canonical ensemble were identified using the MDAnalysis Python library\cite{Gowers2016,Michaud‐Agrawal2011} (version 2.1.0).
Hydrogen bonds were defined geometrically with a cutoff distance of 3.5 \si{\angstrom} and a cutoff angle of 150°.
Water clusters were then selected based on connectivity across all configurations.
For instance, a pair of hydrogen-bonded molecules with indices {0,1} in configuration at $t=0$ and another pair with indices {1,2} at $t=50$ constitute a cluster with indices {0,1,2}.
Water molecules in a cluster consisting of three or more water molecules were classified as a clustered molecule, and remaining molecules were classified as isolated.

\section{Results \& Discussion}

\subsection{(Anti)plasticization of Thermomechanical Properties}

We first assess  whether simulations capture the composition-dependent (anti)plasticization of chitosan-water systems, as reflected in the elastic modulus and glass transition temperatures.
Both $E$ and $T_\text{g}$ are common indicators of (anti)plasticization behavior, and $E$ directly relates to material's functional performance.

    \begin{figure}[htb]
\centering
\includegraphics[scale=1.0]{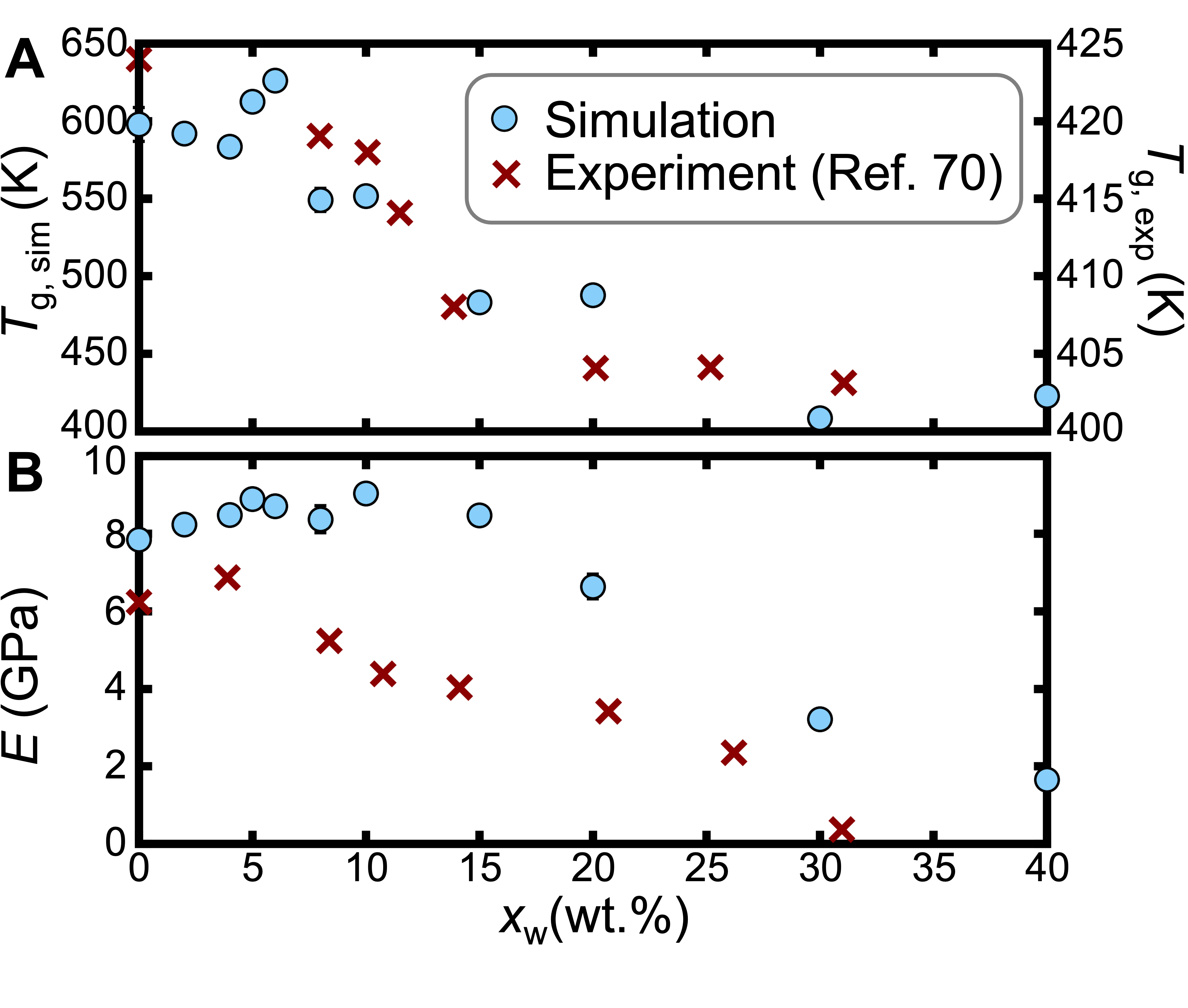}
\caption{Composition dependence of thermomechanical properties.
(A) Comparison of composition-dependent simulated effective $T_\text{g}$ (blue circles) to experimental\cite{AguirreLoredo2016} $T_\text{g}$ (red crosses).
The simulated and experimental values are denoted by the left and right y-axes.
(B) Comparison of composition-dependent simulated $E$ (blue circles) to experimental $E$ (red crosses).
Error bars (black) for simulation data represent the standard error of the mean from four independent simulations and are generally smaller than the markers.}
\label{fig1}
\end{figure}

\autoref{fig1} compares experimental\cite{AguirreLoredo2016} and simulation trends for $x_\text{w}$ = 0, 2, 4, 5, 6, 8, 10, 15, 20, 30, 40\% $T_\text{g}$ and $E$, the Young's modulus.
Compared to neat polymer, the simulated system \rev{at 5 and 6 wt.\%} water displays a higher $T_\text{g}$ as shown in \autoref{fig1}A.
Above this range, the material undergoes rapid plasticization with a drop in $T_\text{g}$ that plateaus above 30 wt.\% water.
The simulated $T_\text{g}$ values are systematically higher than experiment and exhibit a much larger magnitude of change.\cite{Chu2020,Soldera2006,Webb2015,Lyulin2003}
\rev{This discrepancy may, in part, be attributable to the substantially faster cooling rates employed in the simulations.}\cite{Hung2020,Moynihan1974}
While rapid plasticization trends are generally in agreement with experiment, their comparison remains inconclusive due to the lack of data at ~5 wt.\%.
Furthermore, the magnitude of $T_\text{g}$ indicates that the material is highly arrested with very little relaxation, making the changes in $T_\text{g}$ difficult to resolve.

\autoref{fig1}B displays the variation of Young's moduli with increasing water content.
The simulated moduli reasonably agree with both the magnitude and water-content dependence of the experimental values.
The material exhibits a notable increase in $E$ from dry polymer to \rev{$<$10 wt.\% water}, indicative of antiplasticization.
This is followed by a reduction in $E$ at higher water contents, with a more pronounced decrease above 15 wt.\% water, suggesting a crossover to plasticization.
The simulated $E$ values are systematically higher than experiment likely due to higher strain rate associated with simulations.
However, the consistency of trends suggests that the underlying physics of (anti)plasticization behavior is captured by simulations.
\rev{While these results provide clear evidence of (anti)plasticization behavior, the precise composition of the crossover between the two regimes is not resolved by the present data and is beyond the scope of this work.}
Additional results for the bulk modulus, shear modulus, and the Poisson's ratio are included in Supplemental Information Figure S2.
These mechanical properties generally follow the same trends as the elastic modulus, with the exception of the bulk modulus, which exhibits a delayed transition from antiplasticization to plasticization.

\subsection{Decomposition of Elastic Moduli Contributions}
\begin{figure}[h]
\centering
\includegraphics[scale=0.9]{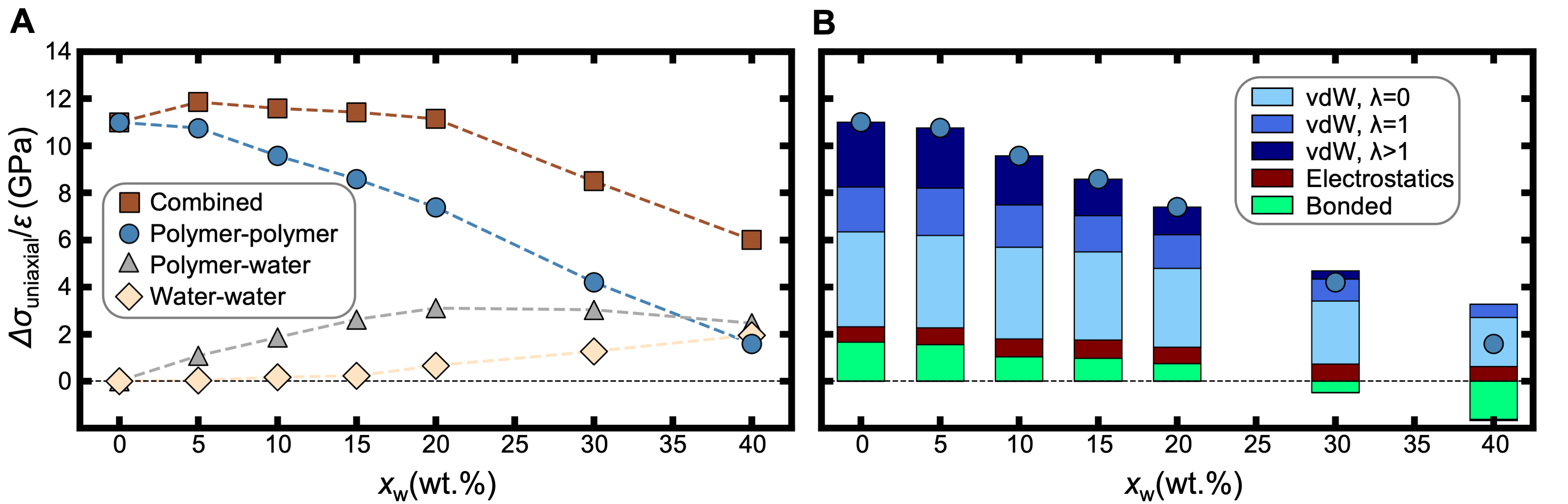}
\caption{Analysis of species-level contributions to Young's modulus trends.
(A) Contributions to elastic moduli from interactions between all (dark red), polymer-polymer (blue), polymer-water (gray), and water-water (beige) species are displayed as a function of water content. Dashed lines connect markers only as guides to the eye.
(B) Contributions to elastic moduli from polymer-polymer interactions are further decomposed into intra-monomer (light blue), neighboring monomer (blue), remote (dark blue) van der Waals interactions.
Electrostatic (brown) and bonded (green) contributions represent the sum from all polymer-polymer interactions.}
% Error bars (black) for simulation data, only displayed in (A), represent the standard error of the mean from four independent simulations and are generally smaller than the markers.
\label{fig2}
\end{figure}

Having established that our model captures experimental sensitivity to changes in water content, we now turn to understand what underlies the mechanical (anti)plasticization of chitosan by water.
In particular, we aim to disentangle the balance of intermolecular interactions by attributing the observed trends in $E$ to specific interactions, thereby providing mechanistic insight into their origin.
We decompose the elastic modulus into its contributing species (polymer-polymer, polymer-water, water-water) and interactions (bonded, electrostatic, van der Waals) in \autoref{fig2}.
Using Eq. {\eqref{stress}} allows for attribution of these trends by computing stress from the derivative of the total energy with respect to volume, which can be decomposed into contributions from individual species and interaction types.

\autoref{fig2}A depicts the contribution of each component to the overall elastic response across the composition range at the species population level.
In the antiplasticization regime, as water content increases, so too does the stress associated with polymer-water interactions.
By contrast, change in stress associated with polymer-polymer interactions is roughly constant.
We interpret this to mean that water-polymer interactions directly underlie the increase in modulus associated with antiplasticization.
As water content increases into the plasticization regime, however, the contribution to stress associated with polymer-water interactions plateaus at around 20 wt.\% water, whereas polymer-polymer interactions display a steady decrease.
This generally signifies weaker cohesion between polymer chains as water pervades the system.
This contrasts with mechanisms directly attributing plasticization to polymer-additive interactions,\cite{Stukalin2010,Chen2018} however, as polymer-polymer interactions are predominantly those varying while water content increases in this regime. 

To better understand how polymer-polymer interactions are weakened in the plasticization regime, we decompose the polymer-polymer contributions based on interaction types (bonded and nonbonded) in \autoref{fig2}B.
Nonbonded interactions are further categorized as electrostatic  and van der Waals interactions within each monomer (contour distance, $\lambda=0$), between neighboring monomers ($\lambda=1$), and between monomers separated by at least one monomer unit ($\lambda>1$).
By this analysis, van der Waals interactions across all groups (blues) dominate the changes in modulus across the composition range.
Interactions within the same monomer (light blue) and neighboring monomer units (blue) notably contribute.
While remote interactions (dark blue) have a considerable impact on the changes in moduli, they do not solely account for the plasticization trends above 10 wt.\% water.
The significant contributions from intra- and neighboring monomer interactions suggest that glucosamine ring conformation and consequent intra-monomer interactions may play a role in the (anti)plasticization phenomena.
Electrostatic interactions minimally impact the modulus with a very small dependence on the water content.
While bonded interactions become increasingly plasticizing with increasing water content, the magnitude of observed changes remains limited.

    \begin{figure}[htb!]
\centering
\includegraphics{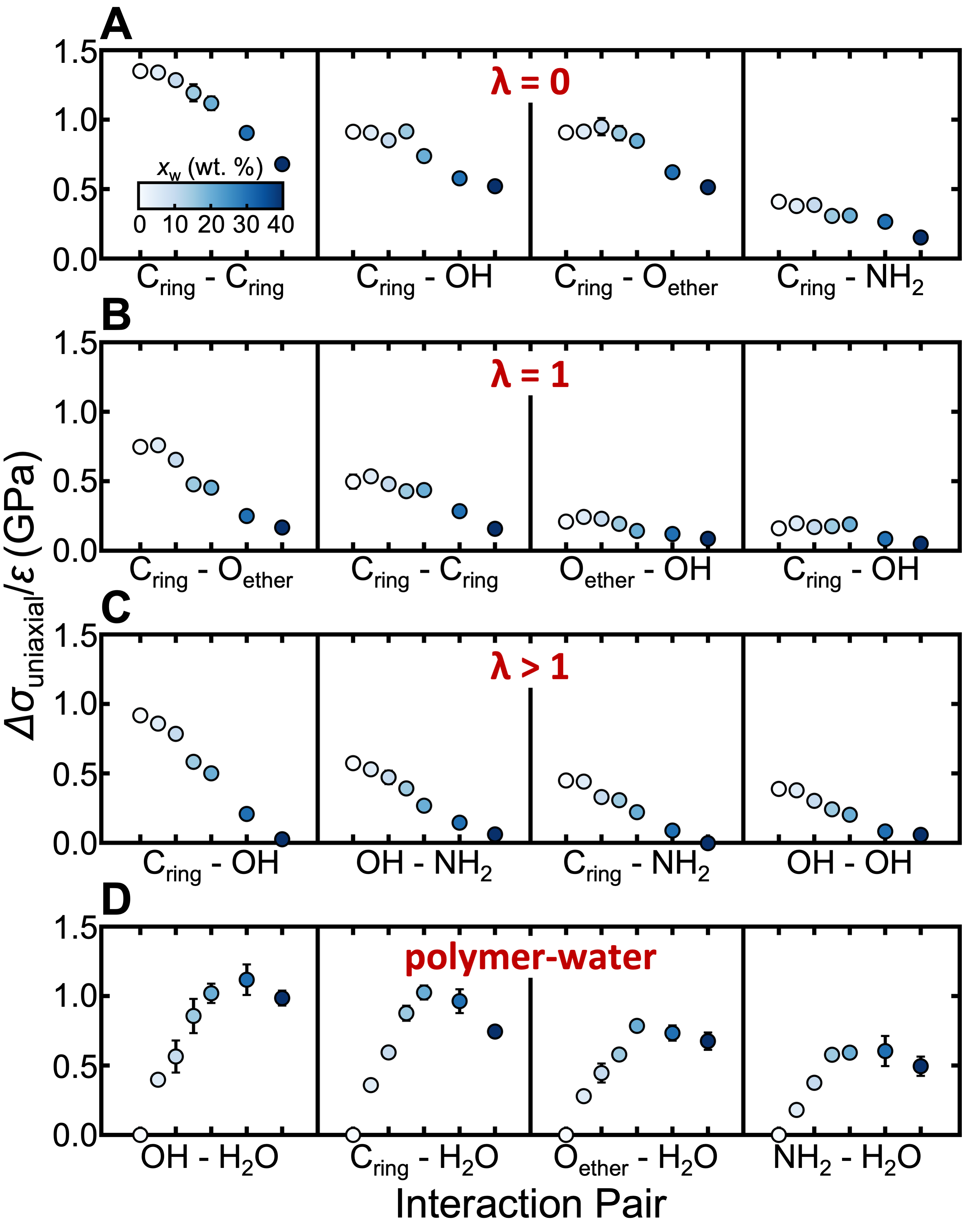}
\caption{Contributions to Young's moduli from specific interactions between pairs of chemical groups.
Sum of contributions between the top four interaction pairs for (A) intra-monomer, (B) neighboring monomer, (C) remote polymer-polymer interactions, and (D) polymer-water interactions are displayed.
The interaction pairs are sorted based on the respective magnitude of contributions at 0 wt.\% water for polymer-polymer interactions (A, B, C); at 5 wt \% water for polymer-water interactions (D). 
Analyzed chemical groups are categorized as backbone alkane carbon and hydrogens (C$_\text{ring}$), ether oxygen (O$_\text{ether}$), hydroxyl (OH), and amine (NH$_\text{2}$) species.
The color bar and x-axis ticks in (A), which denote the water content, applies to all axes.
Error bars (black) for simulation data represent the standard error of the mean from four independent simulations and are generally smaller than the symbol size.
Extended analysis containing normalized interaction contributions are displayed in Supplemental Information, Figure S6}
\label{fig3}
\end{figure}

To understand the nanoscale origins of van der Waals contributions, we characterize each interaction based on participating chemical groups.
Decomposition of intra-monomer interactions in \autoref{fig3}A indicates that all four dominant contributions involve a backbone carbon, with C$_\text{ring}$–C$_\text{ring}$ interactions contributing most prominently across the composition range.
Based on these trends, glucosamine rings do not rearrange to a stable conformation upon deformation, imparting additional stress that increases the modulus. 
Generally, intra-monomer contributions between the four pairs remain constant at low water contents, with C$_\text{ring}$-C$_\text{ring}$ and C$_\text{ring}$-OH pairs showing a small initial increase.
As such, water molecules at low contents do not facilitate glucosamine ring relaxation and instead have an antiplasticizing effect.
Contributions from neighboring monomers in \autoref{fig3}B are broadly smaller than those from intra-monomer interactions.
Nevertheless, they also include backbone carbons, with C$_\text{ring}$-O$_\text{ether}$ and C$_\text{ring}$-C$_\text{ring}$ interactions as the dominant contributors.

Contributions of interactions between monomers separated by at least one monomer unit ($\lambda>1$) in \autoref{fig3}C reveal different trends.
Sidechains containing OH and NH$_2$ groups dominate the interactions in this group, in contrast to closer-range interactions, highlighting a shift in impact as the interaction distance increases.
The significant contributions from OH-NH$_2$ and OH-OH interactions can be attributed to polymer-polymer hydrogen bonds.
However, major contributions from C$_\text{ring}$-OH and C$_\text{ring}$-NH$_2$ pairs indicate that weaker nonbonded interactions considerably impact the elastic moduli.
Upon deformation, side groups are unable to rearrange into regions unoccupied by other polymer chains, resulting in increased stress caused by interactions with the polymer backbone.
In contrast with intra- and neighboring monomer interactions, far-range contributions monotonically decrease with water content across the composition range for all interaction pairs.
Far-range interactions, including those between two chains, are strictly plasticized by water molecules, even at low water contents.

The relative stability of intra- and neighboring interaction contributions below 10 wt.\% water alone does not explain the pronounced antiplasticization effect on the material moduli.
Decomposition of polymer-water interaction contributions in \autoref{fig3}D reveals the origins of antiplasticization at low water contents.
Water interactions with polymer hydroxyl, amine, and ether groups highlight the prevalent hydrogen-bonding interactions.
However, interactions between water and backbone carbons, despite their weak electrostatic nature, contribute significantly to the elastic modulus.
Across all polymer-water interaction pairs, the contributions to the elastic moduli reach a peak at 20-30 wt.\% water content.
This observation is in line with the saturation of available polymer groups and material density (Supplemental Information Figure S3 and S4), where all polymer side chains are occupied by the interactions with water molecules.

   \begin{figure}[htb!]
\centering
\includegraphics[scale=0.9]{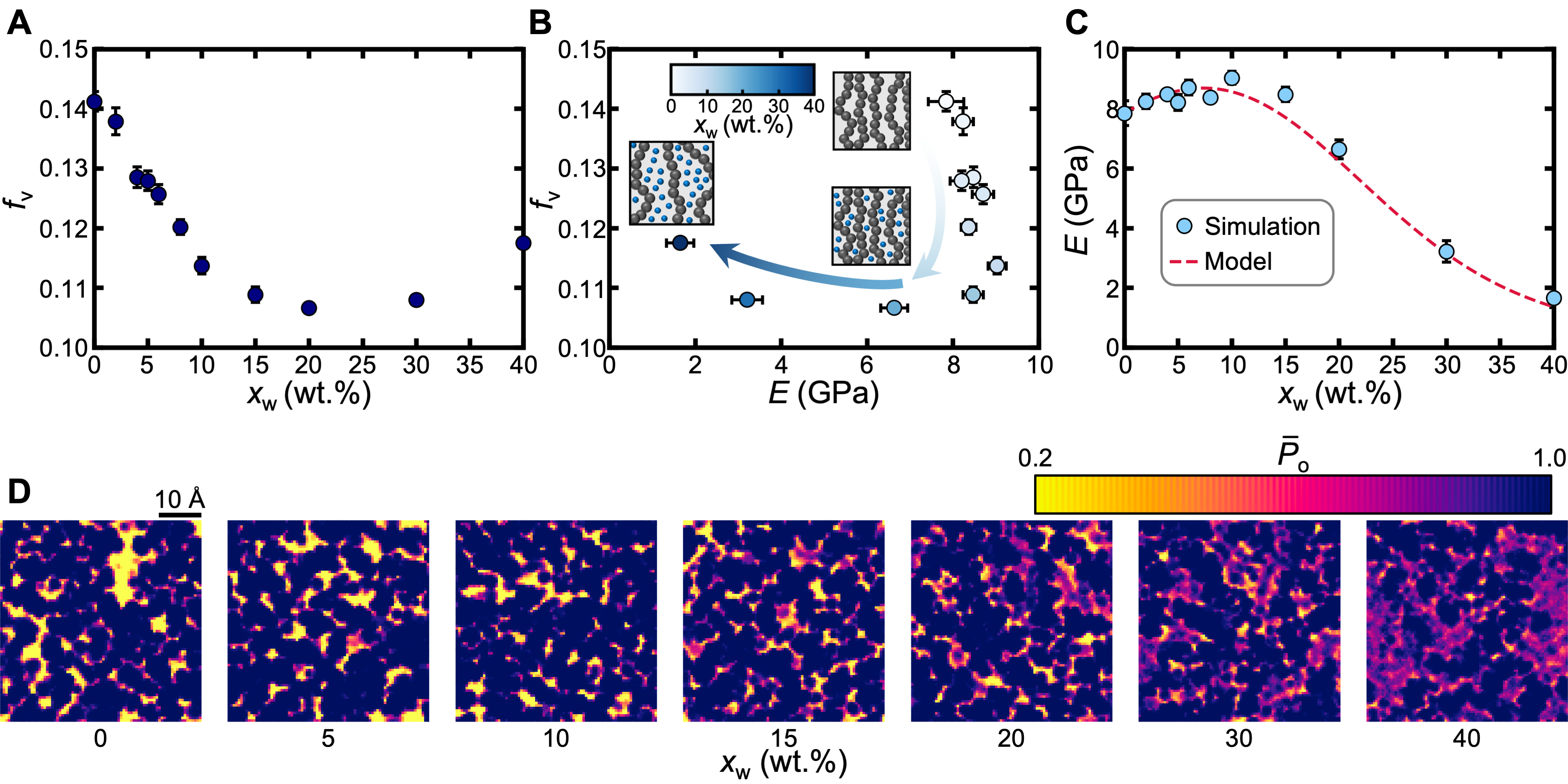}
\caption{Overview of free volume trends.
(A) Composition dependence of the free-volume fraction $f_\text{v}$ for simulated chitosan-water mixtures.
(B) Direct comparison of $E$ and $f_\text{v}$ across the composition range.
The marker colors denote the water content indicated by the color bar.
The illustrations visualize the hypothesized mechanisms underlying free volume trends, where polymer and water species are displayed in dark gray and blue.
The background (light gray) represents the free volume regions.
(C) Comparison of simulated elastic modulus values with the predictions from the developed model (Eq. \eqref{eq:modulus_model}).
The blue markers represent the measured elastic modulus, and the dashed line indicates the model fit.
For the model plot, both $f_\text{v}$ and $\bar{p}_\text{o,free}$ are interpolated using a quartic function to obtain a smooth fit containing 100 data points across the composition range, and the resulting fit is shown in Supplemental Information Figures S5 and S6.
Error bars (black) represent the standard error of the mean from four independent simulations and are not visible behind the markers.
(D) Simulation snapshots visualizing the discretized grid points with their respective occupancy probabilities $\bar{p}_\text{o}$ for all investigated systems. The color bar indicates the occupancy probability of a grid point.
Only grid points with the lowest z-value in the mesh are shown for visual clarity.
All simulation snapshots were generated using the Visual Molecular Dynamics (VMD) software (version 1.9.4)\cite{HUMP96}
}
\label{FV}
\end{figure}

\subsection{Assessment of Free Volume Effects}\label{S:structuralanalysis}
The prior analysis featuring decomposition of elastic moduli contributions highlighted competing effects between weakened polymer-polymer and enhanced polymer-water interactions, with the former seemingly dominating the plasticization regime.
However, the prior decomposition cannot resolve water-mediated structural perturbations that may govern the decreasing contributions from polymer-polymer interactions.
Such mediated contributions may arise, for example, by water increasing inter-chain spacing or by imparting additional accessible free volume.
To gain some insight into such effects, we examine how free volume varies with composition and correlates with the elastic modulus (\autoref{FV}).

\autoref{FV}A indicates that adding water to polymer-rich systems ($<$ 20 wt.\%) decreases the fraction of free volume, but continuing to add water results in a moderate increase in the fraction of free volume.
The reduction in polymer-rich systems is attributed to water initially filling interchain voids.
However, changes in free volume are not directly in line with the mechanical (anti)plasticization trends observed  in \autoref{fig1}.
This is demonstrated by direct comparison in \autoref{FV}B, in which
the two properties display a nonmonotonic relationship across the composition range.
While $f_\text{v}$ rapidly decreases at low water contents, its impact on $E$ remains limited below 15 wt.\%.
Above 20 wt.\%, however, $E$ is highly sensitive to water content during rapid plasticization in contrast with a subtle increase in $f_\text{v}$.
Consequently, both the fraction of free volume and elastic moduli display distinct behavior at low- and high content regimes, and the modulus cannot be uniquely determined by specification of the free-volume fraction. 

\subsection{Persistent Free Volume Model for the Elastic Modulus}

The inability to cast $E$ as a function of $f_\mathrm{v}$ stems from the nature of our calculated free volume, as illustrated by the insets of \autoref{FV}B.
In particular, the static free volume fraction $f_\mathrm{v}$ provides a geometric measure of unoccupied 
space but does not distinguish between free-volume regions of different dynamical character. Regions occupied by mobile water molecules may be readily viewed as unoccupied free volume due to diffusion and fluctuations, and likewise unoccupied voids may be transiently visited by such mobile species. 
White and Lipson draw an important distinction between total free volume and excess free volume, where the latter excludes contributions from the vibrational volume $V_\mathrm{vib}$ associated with solid-like segmental motion.\cite{White2016} 
In their framework, the excess free volume enables cooperative molecular rearrangement, while the vibrational contribution is mechanically inert. 
The geometric $f_\mathrm{v}$ computed is more analogous in spirit to a total free volume, counting all instantaneously unoccupied space without regard to whether those voids are  genuinely persistent.
Here, we expect that persistent, genuinely unoccupied voids are those that contribute more to the modulus than those that are dynamically accessible to mobile species, which may move into free-volume voids on timescales faster than the deformation response.

To account for this, we introduce a time-averaged occupation probability $\bar{p}_\mathrm{o}$ (see Eq. \eqref{probability}), which corresponds to the probability that a grid point is instantaneously occupied at any moment during an observation window by a particle, given that it was initially designated as free volume (at $t=0$).
Voids that are frequently sampled by thermally fluctuating atoms have comparatively large $\bar{p}_\mathrm{o}$, while persistently unoccupied voids have small $p_\mathrm{o}$. 
We subsequently define a corrected effective free volume
\begin{equation}
    \tilde{f}_\mathrm{v}(x_\mathrm{w}) = \frac{f_\mathrm{v}(x_\mathrm{w})}
    {\bar{p}_\mathrm{o}(x_\mathrm{w})},
    \label{eq:corrected_fv}
\end{equation}
which amplifies the contribution of persistently unoccupied voids relative to dynamically accessible ones. 
Although it is not mathematically transparent, this is physically resonant with discounting initially excluded volume based on transient fluctuations of mobile species. 

The elastic modulus is then well described by
\begin{equation}
    E(x_\mathrm{w}) = E_0 \left( \frac{\tilde{f}_\mathrm{v}(x_\mathrm{w})}
    {\tilde{f}_{\mathrm{v},0}} \right)^2
    \label{eq:modulus_model}
\end{equation}
where $E_0$ and $\tilde{f}_{\mathrm{v},0}$ denote the Young's modulus and effective free volume of the neat polymer reference state. 
This formulation expresses the modulus as a projection from the reference, such that the prediction depends entirely on how $\tilde{f}_\mathrm{v}$ changes with water content, 
making it more robust to systematic offsets in either $f_\mathrm{v}$ or $\bar{p}_\mathrm{o}$ (since both quantities have associated hyperparameters required for analysis).
The origin of the squared exponent in Eq.~(\ref{eq:modulus_model}) can be physically motivated by the tensorial structure of the elastic constants. 
Specifically, in the Born-Huang formulation, the elastic constant tensor is given by
\begin{equation}
    C_{ijkl} = \frac{1}{V} \sum_{\alpha} \sum_{\beta \neq \alpha} 
    \frac{\partial^2 U}{\partial \mathbf{r}_{\alpha}[j] \, \partial \mathbf{r}_{\beta}[k]} \, 
    \mathbf{r}_{\alpha\beta}[j] \, \mathbf{r}_{\alpha\beta}[l]
    \label{eq:born_huang}
\end{equation}
where $U$ is the total potential energy, $\mathbf{r}_{\alpha}[i]$ denotes the $i$-th Cartesian component of the position of atom $\alpha$, $\mathbf{r}_{\alpha\beta}[j]$ is the $j$-th component of the separation vector between atoms $\alpha$ and $\beta$, and $V$ is the system volume. 
This demonstrates how elastic constants are fundamentally pairwise quantities, as stress transmission requires two sequentially connected load-bearing sites. 
Supposing that $\tilde{f}_\mathrm{v} / \tilde{f}_{\mathrm{v},0}$ characterizes the fractional change in the density of mechanically active sites relative to the reference, the density of load-bearing pairs scales as the square of this quantity, and the elastic modulus inherits this dependence.
In the static limit, if $\bar{p}_\mathrm{o}$ and $\bar{p}_{\mathrm{o},0}$ both approach zero at the same rate, Eq.~(\ref{eq:modulus_model}) reduces to a purely geometric scaling $E = E_0(f_\mathrm{v}/f_{\mathrm{v},0})^2$. 
In the high-mobility limit, $\bar{p}_\mathrm{o}$ approaches a finite value set by the local instantaneous particle density near free-volume regions.
Thus, the model appears well-posed across the physically relevant composition range.
 
\autoref{FV}C shows that this effective dynamical free volume model quantitatively reproduces the composition-dependent 
elastic modulus across the full range of water contents investigated.
It neatly captures both the antiplasticization peak and the subsequent rapid decrease associated with plasticization. 
Notably, the model achieves this with no freely adjustable parameters beyond analytical fits to $f_\mathrm{v}(x_\mathrm{w})$ 
and $\bar{p}_\mathrm{o}(x_\mathrm{w})$, both of which are computed directly from the simulation trajectories used for the mechanical analysis and behave predictably.
\rev{Furthermore, the model does not require a specific crossover concentration for (anti)plasticization as it applies across both regimes as a continuous function of composition.}
\autoref{FV}D provides qualitative support for the physical picture underlying the model by illustrating the spatial distribution of occupancy probabilities $\bar{p}_\mathrm{o}$ across the composition range. 
At low water contents, free-volume grid points are characterized by broadly elevated $\bar{p}_\mathrm{o}$, reflecting frequent dynamical visitation by thermally fluctuating polymer atoms and the immobile water molecules that occupy interchain voids without imparting persistent free volume. 
At higher water contents, a growing population of grid points with low $\bar{p}_\mathrm{o}$ emerges, consistent with the formation of bulk water domains that introduce genuinely persistent, mechanically relevant free volume and drive the transition to plasticization.

    \begin{figure}[htb]
\centering
\includegraphics[scale=0.9]{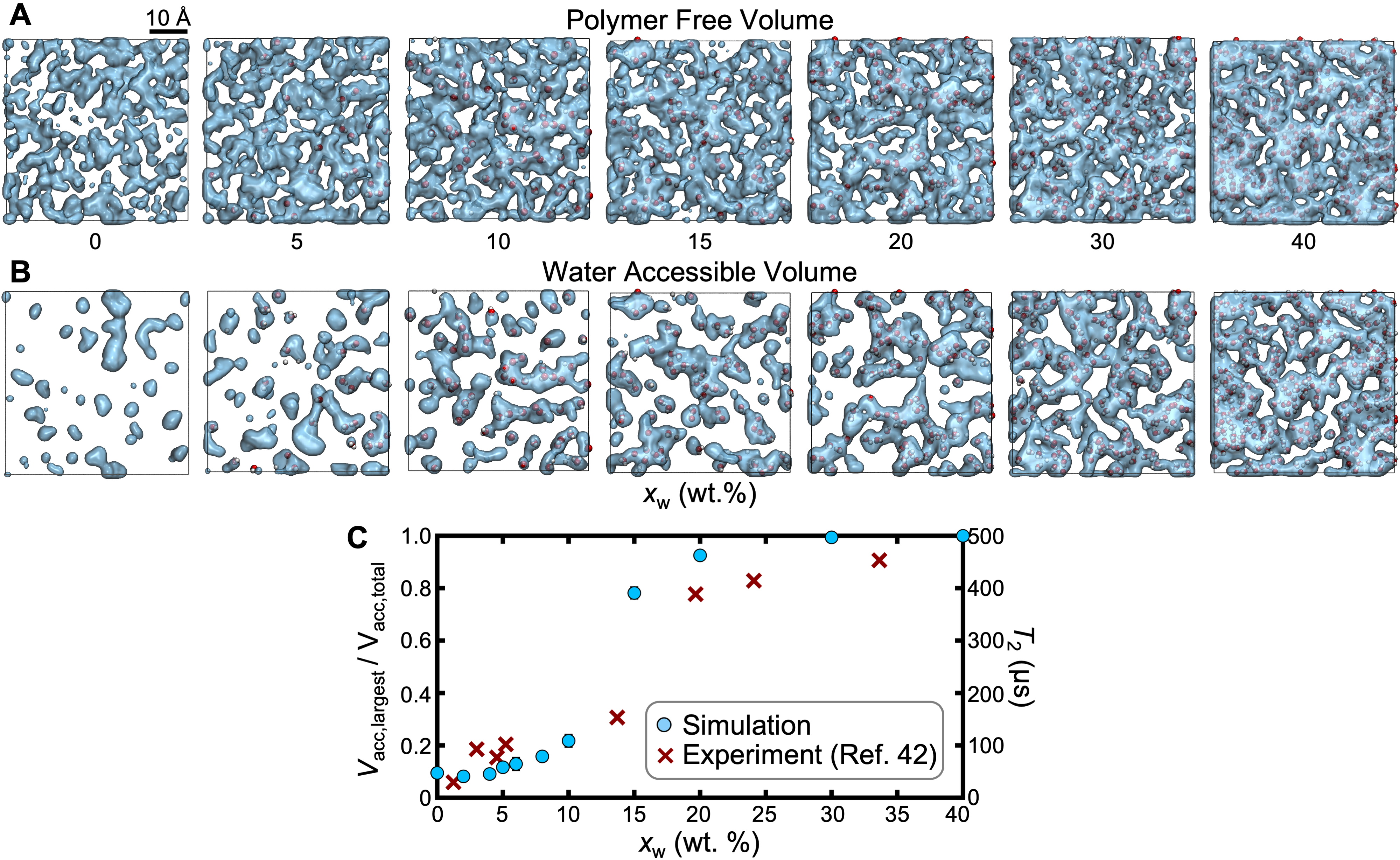}
\caption{Summary of free volume trends.
(A) Molecular renderings showing the free volume regions (blue) unoccupied by polymer atoms across the composition range, with water molecules explicitly displayed.
(B) Free volume regions (blue) unoccupied by polymer atoms and can accessed by water molecules, also including explicit water molecules.
Regions and atoms with $z$ coordinates below 5\si{\angstrom} are displayed to highlight connectivity.
Oxygen and hydrogen atoms in (A) and (B) are explicitly visualized in red and white.
(C) Fraction of largest water-accessible volume cluster to the total water-accessible free volume compared with experimental\cite{MadeleinePerdrillat2016} spin-spin relaxation times $T_2$ across the composition range.
Error bars (black) in (C) represent the standard error of the mean from four independent simulations.
All molecular renderings were generated using the Visual Molecular Dynamics (VMD) software (version 1.9.4)\cite{HUMP96}
}
\label{WAV}
\end{figure}

\subsection{Water-Accessible Volume as a Key Driver of Plasticization}

Next, we examine what underlies the persistence or dynamic accessibility of a free volume grid point within the system.
 \autoref{WAV}A illustrates the connectivity of polymer-unoccupied voids ($V^{\mathrm{polymer}}_\mathrm{free}$) with explicit water embedded within.
While $V^{\mathrm{polymer}}_\mathrm{free}$ clearly increases with water content, the voids are fully connected across the entire composition range.
This total percolation behavior is attributed to narrow intermolecular channels that cannot be accessed by water molecules.
To address this, we further refine our analysis to compute a so-called water-accessible volume, $V_\mathrm{acc}$.
\autoref{WAV}B reveals that water-accessible voids are isolated and scattered throughout the polymer-rich systems.
At 10-15 wt.\% water, these voids begin to connect.
Above 20 wt.\%, all voids are effectively fully connected, and water molecules can fully percolate and traverse the polymer matrix.
\autoref{WAV}C offers mechanistic insight into how  the connectivity of water-accessible voids relate to a crossover from antiplasticizing to plasticizing behavior.
In particular, the percolation of voids, probed by ratio of the largest contiguous water-accessible volume to all water-accessible volume, coincides with when the elastic modulus begins to notably decrease. 
These trends also neatly track with experimental low-field NMR data probing water relaxation, which show a marked increase in spin-spin relaxation dynamics between 15-20 wt.\% water.\cite{MadeleinePerdrillat2016}
The spin-spin relaxation time $T_2$ is sensitive to the translational mobility of water molecules, with longer $T_2$ values reflecting less restricted, more bulk-like dynamics. 
This corroborates the view  in \autoref{WAV}B,C, of a transition from more isolated water clusters to a percolated network in which water can exchange freely across the matrix, and this percolation promotes plasticization.

\begin{figure}[htb]
\centering
\includegraphics[scale=1.0]{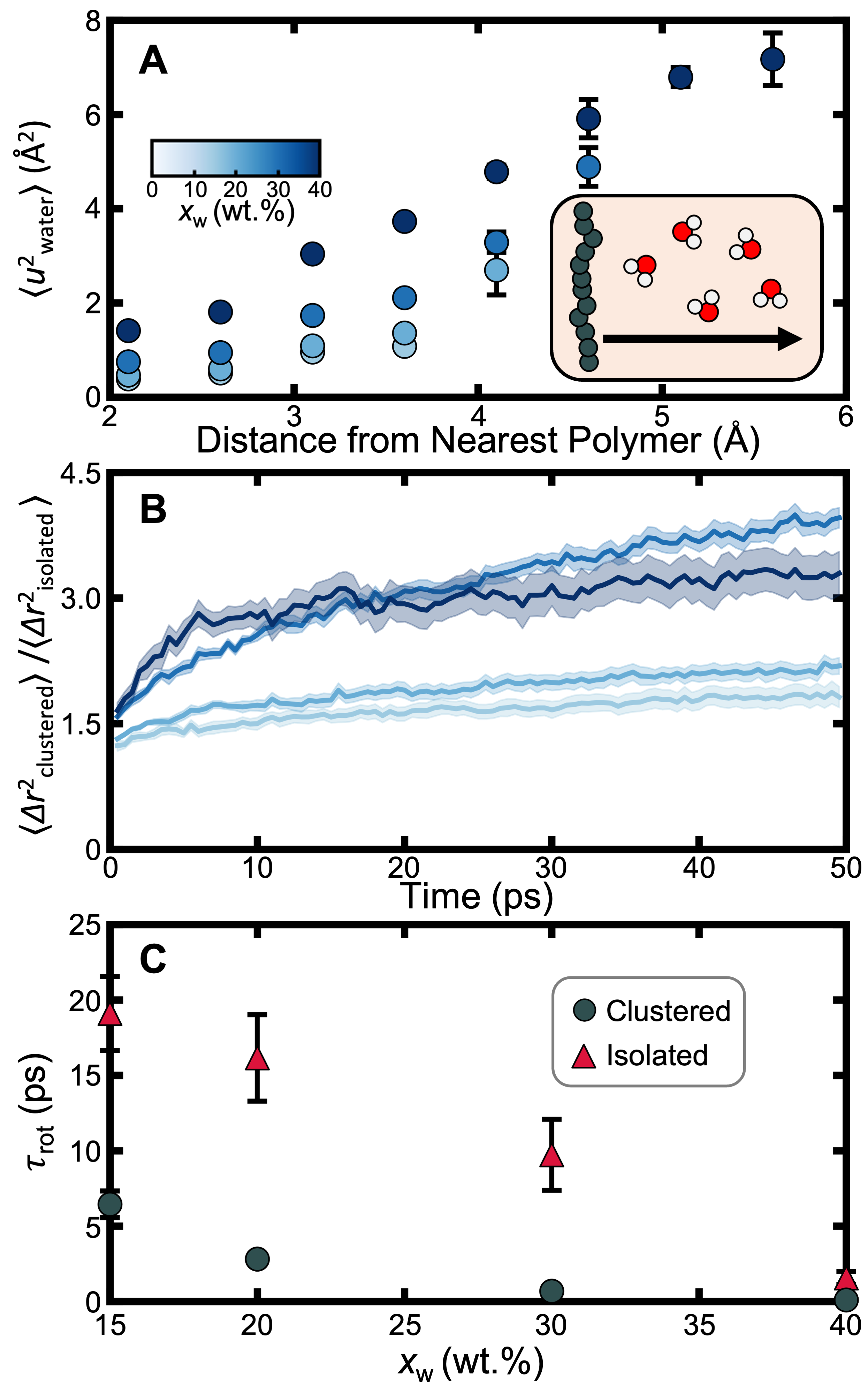}
\caption{Dynamic heterogeneity of water in chitosan-water mixtures.
(A) Vibrational mobility of water molecules, measured by the Debye-Waller factor, as a function of distance from nearest polymer heavy atom.
The distances were discretized into 0.5 \si{\angstrom} bins.
The marker colors denote the water content indicated by the color bar.
(B) Comparison of translational mobility, measured by the mean-squared displacement (MSD), of clustered and isolated water molecules over a 50 ps range.
(C) Comparison of orientational mobility, measured by the rotational relaxation time constants, of cluster (gray marker) and isolated (red marker) water molecules.
Only systems with water contents of 15, 20, 30, and 40 wt.\% are displayed in (A), (B), and (C) due to lack of clustered molecules at lower water contents.
The criteria for cluster and isolated water molecules are defined based on participation in a hydrogen-bonding network comprising three or more molecules.
The error bars in (A) and (C), and the error regions in (B) represent the standard error of the mean from four independent simulations.}
\label{fig5}
\end{figure}

\subsection{Effect of Local Environment on Additive Dynamics}

Given the alignment with the $T_2$ relaxation trends, we sought to further characterize the water dynamics. \autoref{fig5}A shows that water mobility increases substantially with distance from polymer chains across all compositions, but the gradient is most pronounced above 20~wt.\% water, where bulk-like water domains form far from the polymer chains. 
Meanwhile, \autoref{fig5}B shows that clustered water molecules, defined by participation in a hydrogen-bonding network of three or more molecules, exhibit markedly higher translational mobility than isolated molecules across all compositions investigated.
The ratio of the apparent diffusivity, which could be estimated from the mean squared displacement, is about three for the higher water content systems.
Finally, \autoref{fig5}C compares the rotational water relaxation dynamics, for which isolated molecules at low water contents exhibit relaxation time constants exceeding 15~ps, whereas clustered molecules relax around an order of magnitude faster. At higher water contents, the isolated population is effectively absent as nearly all water molecules participate in percolating clusters, and the distinction between the two populations vanishes. 
Together, these results demonstrate that there are dynamically distinct populations of water molecules that persist well into the plasticization regime, with some motionally restricted and isolated and others more active far from the polymer chains. 

\section{Conclusions}
This study elucidates the nanoscale origins of (anti)plasticization in chitosan-water mixtures. At low water contents, effectively immobile water molecules fill interchain voids and interact directly with polymer functional groups, resulting in an increase in the elastic modulus. 
At higher water contents, percolation of water clusters throughout the polymer matrix enables cooperative molecular motion, reducing interchain interactions and enhancing conformational relaxation of glucosamine rings, driving the crossover to plasticization. 
To capture the competition between these mechanisms across the full composition range, we introduce a persistent free volume model, which incorporates both the fraction and the dynamical accessibility of free-volume regions, which quantitatively reproduces the 
antiplasticization peak and subsequent rapid decrease in modulus without freely adjustable parameters. 
The model reflects a physically transparent picture in which persistent, genuinely unoccupied voids sustain elastic stress while voids sampled by mobile water molecules do not, such that the dynamic accessibility of free volume governs the crossover from antiplasticization to plasticization. These insights may be generalizable to other bio- and synthetic polymers and help inform the design of polymer-additive systems.

While our findings broadly agree with lubricity and free-volume theories, neither framework alone sufficiently elucidates the (anti)plasticization of elastic moduli across the full composition range. 
Overall, species-level interaction contributions partially align with lubricity 
theory, where reduced polymer-polymer interactions at higher water contents result in net plasticization, and the enhanced vibrational mobility of clustered water molecules away from the polymer surface is consistent with the formation of gliding planes.
Although the gliding planes cannot be discerned on the timescales of our simulations, we do identify a residual population of motionally restricted, isolated water molecules that persists well into the plasticization regime, demonstrating that complete homogenization of additive dynamics is not a prerequisite for the onset of plasticization.
However, significant contributions to $E$ from intra-monomer ($\lambda = 0$) and neighboring monomer ($\lambda = 1$) interactions, including conformational relaxation of glucosamine rings, are not captured by lubricity theory, which primarily considers interchain friction ($\lambda > 1$). 
Similarly, the decrease in modulus at high water contents is broadly consistent with free-volume theory, as is the antiplasticizing role of polymer-water interactions including both hydrogen bonds with hydroxyl and amine groups and weaker van der Waals contacts with backbone carbons. 
However, the relationship between free volume fraction and modulus is non-monotonic across the composition range, and free volume decreases at the same low water contents where antiplasticization is observed. 
We show that it is the persistent, dynamically stable fraction of free volume, rather than the total geometric free volume, that governs the elastic response. 
The persistent free volume model incorporates this distinction and conceptually extends the idea of free-volume theory to quantitatively captures the (anti)plasticization behavior across the full composition range.

Several limitations of the present study highlight opportunities for future directions. 
The specific observations and analyses are specific to chitosan-water mixtures, such that it will be of interest to evaluate the generalizability of the identified mechanisms in other biopolymer-additive 
systems.
\rev{In addition, a systematic investigation of molecular weight dependence would further validate our conclusions and help clarify how chain length influences free-volume persistence and (anti)plasticization behavior.}
\rev{The molecular weights evaluated in this study are below the regime where significant entanglement effects have been reported for chitosan.\cite{Seggern2025}
Consequently, the mechanisms identified in this work are expected to be governed primarily by local intermolecular interactions, chain packing, and free-volume accessibility rather than chain-scale dynamical processes such as constraint release or reptation.}
\rev{Another consideration is the role of composition-dependent amine protonation.
The intrinsic pKa of polymeric chitosan amines ($\sim$6.3--6.8) implies that a substantial fraction of amines remain unprotonated, and the system resides largely in the neutral regime.\cite{Morrow2015,Wang2006}
While our simulations are broadly consistent with these near-neutral conditions relevant to most experimental systems, protonation may influence properties by modulating intermolecular interactions, water uptake, and chain packing.
Although we do not expect this to qualitatively alter the presented mechanistic framework of (anti)plasticization, future studies incorporating composition-dependent charge regulation, such as via constant-pH molecular dynamics,\cite{MartinsdeOliveira2022,Morrow2015} would provide a more rigorous assessment.}
The persistent free volume model may also be better tested in systems with lower $T_\mathrm{g}$, where thermomechanical properties in the polymer-rich regime are more tractable to resolve computationally.
Future work expanding the simulation framework to diverse polymer-additive systems could also probe how backbone flexibility, side-chain chemistry, and additive mobility govern the dynamical accessibility of free-volume regions.
This may help establish chemical guidelines on the (anti)plasticizing effect of different additives across composition ranges.

\FloatBarrier

\begin{acknowledgement}
 B.E.U. and M.A.W. acknowledge support from the National Science Foundation under Grant No. 2237470. 
 Simulations and analyses were performed using resources from Princeton Research Computing at Princeton University, which is a consortium led by the Princeton Institute for Computational Science and Engineering (PICSciE) and Office of Information Technology's Research Computing.
 These resources include a GPU-based computing cluster purchased with support from the National Science Foundation (Grant No. NSF-MRI: OAC-2320649) (M.A.W)
\end{acknowledgement}

\section*{Data Availability}
All molecular dynamics simulation input files are publicly available via GitHub at https://github.com/webbtheosim/md-simulation-files/tree/main/2026-chitosan-water.

\section*{Supplemental Information}
Simulation Settings. Glass Transition Temperature Analysis. Analysis of Additional Mechanical Properties. Saturation of Polymer Groups. Composition Dependence of Material Density. Occupancy Probability Analysis. Normalized Contributions to Young's Moduli.

\bibliography{references}

\end{document}